\input harvmac
\overfullrule=0pt
%
\def\simge{\mathrel{%
   \rlap{\raise 0.511ex \hbox{$>$}}{\lower 0.511ex \hbox{$\sim$}}}}
\def\simle{\mathrel{
   \rlap{\raise 0.511ex \hbox{$<$}}{\lower 0.511ex \hbox{$\sim$}}}}
 
\def\slashchar#1{\setbox0=\hbox{$#1$}           
   \dimen0=\wd0                                 
   \setbox1=\hbox{/} \dimen1=\wd1               
   \ifdim\dimen0>\dimen1                        
      \rlap{\hbox to \dimen0{\hfil/\hfil}}      
      #1                                        
   \else                                        
      \rlap{\hbox to \dimen1{\hfil$#1$\hfil}}   
      /                                         
   \fi}                                         %
\def\ts{\thinspace}
\def\tx{\textstyle}
\def\ra{\rightarrow}

\def\ol{\bar}

\def\CC{{\cal C}}
\def\CD{{\cal D}}

\def\CI{{\cal I}}

\def\CN{{\cal N}}
\def\CO{{\cal O}}

\def\CS{{\cal S}}

\def\ecm{\sqrt{s}}
\def\shat{\hat s}
\def\that{\hat t}
\def\uhat{\hat u}
\def\rshat{\sqrt{\shat}}

\def\atc{\alpha_{TC}}
\def\aqcd{\alpha_{S}}
\def\atro{\alpha_{\rho_T}}

\def\Ntc{N_{TC}}
\def\suc{SU(3)}

\def\sutc{SU(\Ntc)}

\def\thw{\theta_W}
\def\kslash{\raise.15ex\hbox{/}\kern-.57em k}
\def\LTC{\Lambda_{TC}}
\def\LETC{\Lambda_{ETC}}

\def\CDgg{\CD_{g g}}
\def\CDgrho{\CD_{g\rho_T}}
\def\tro{\rho_{T1}}

\def\troct{\rho_{T8}} 
\def\tropm{\rho_{T1}^\pm}

\def\troz{\rho_{T1}^0}
\def\tpi{\pi_T}
\def\tpipm{\pi_T^\pm}
\def\tpimp{\pi_T^\mp}
\def\tpip{\pi_T^+}
\def\tpim{\pi_T^-}
\def\tpiz{\pi_T^0}
\def\tpipr{\pi_T^{0 \prime}}
\def\etat{\eta_T}

\def\octpipm{\pi_{T8}^\pm}
\def\octpip{\pi_{T8}^+}
\def\octpim{\pi_{T8}^-}
\def\octpiz{\pi_{T8}^0}

\def\tpilq{\pi_{L \ol Q}}

\def\tpiql{\pi_{Q \ol L}}
\def\tpiun{\pi_{U \ol N}}
\def\tpiue{\pi_{U \ol E}}
\def\tpidn{\pi_{D \ol N}}
\def\tpide{\pi_{D \ol E}}

\def\jets{\rm jets}

\def\pbarp{\ol p p}

\def\gev{{\rm GeV}}
\def\tev{{\rm TeV}}

\def\half{{\textstyle{ { 1\over { 2 } }}}}

\def\twothirds{{\textstyle{ { 2\over { 3 } }}}}

\def\myfoot#1#2{{\baselineskip=14.4pt plus 0.3pt\footnote{#1}{#2}}}

\Title{\vbox{\baselineskip12pt\hbox{FERMILAB-CONF-96/297-T}
\hbox{BUHEP-96-33}
\hbox{hep-ph/9609297}}}
{Electroweak and Flavor Dynamics at Hadron Colliders--I}

\smallskip
\centerline{Estia Eichten\myfoot{$^{\dag }$}{eichten@fnal.gov}}
\smallskip\centerline{Fermi National Accelerator Laboratory}
\centerline{P.O.~Box 500 Batavia, IL 60510}
\centerline{and}
\smallskip
\centerline{Kenneth Lane\myfoot{$^{\ddag }$}{lane@buphyc.bu.edu}}
\smallskip\centerline{Department of Physics, Boston University}
\centerline{590 Commonwealth Avenue, Boston, MA 02215}
\vskip .3in

\centerline{\bf Abstract}

This is the first of two reports cataloging the principal signatures of
electroweak and flavor dynamics at $\pbarp$ and $pp$ colliders. Here, we
discuss some of the signatures of dynamical elecroweak and flavor symmetry
breaking. The framework for dynamical symmetry breaking we assume is
technicolor, with a walking coupling $\atc$, and extended technicolor. The
reactions discussed occur mainly at subprocess energies $\rshat \simle
1\,\tev$. They include production of color-singlet and octet technirhos and
their decay into pairs of technipions, longitudinal weak bosons, or jets.
Technipions, in turn, decay predominantly into heavy fermions. This report
will appear in the Proceedings of the 1996 DPF/DPB Summer Study on New
Directions for High Energy Physics (Snowmass 96).

\bigskip

\Date{9/96}

\vfil\eject

\newsec{Introduction}

This is the first of two reports summarizing the major signals for
dynamical electroweak and flavor symmetry breaking in experiments at the
Tevatron Collider and the Large Hadron Collider. The division into two
reports is done solely to accomodate the length requirements imposed on
contributions to the Snowmass~'96 proceedings. In contrast, the motivations
for these studies are clear: We do not know the mechanism of electroweak
symmetry breaking nor the physics underlying flavor and its symmetry
breaking. The dynamical scenarios whose signals we catalog provide an
attractive theoretical alternative to perturbative supersymmetry models. At
the same time, they give experimentalists a set of high-$p_T$ signatures
that challenge heavy-flavor tagging, tracking and calorimetry---detector
subsystems somewhat complementary to those tested by supersymmetry
searches. Finally, many of the most important signs of electroweak and
flavor dynamics have sizable rates and are detected relatively easily in
hadron collider experiments. Extensive searches are underway in both
Tevatron Collider collaborations, CDF and D\O. We hope that these reports
will inspire and help the ATLAS and CMS Collaborations to begin their
studies.

This report lists some of the major signals for dynamical electroweak and
flavor symmetry breaking in experiments at the Tevatron Collider and the
Large Hadron Collider. Section~2 contains a brief overview of technicolor
and extended technicolor. This discussion includes summaries of the main
ideas that have developed over the past decade: walking technicolor,
multiscale technicolor, and topcolor-assisted technicolor. Hadron collider
signals of technicolor involve production of technipions via $\ol q q$
annihilation and $gg$ fusion. These technipions include the longitudinal
weak bosons $W_L$ and $Z_L$ as well as the pseudo-Goldstone bosons $\tpi$
of dynamical symmetry breaking. The $\tpi$ are generally expected to have
Higgs-boson-like couplings to fermions and, therefore, to decay to heavy,
long-lived quarks and leptons.

The subprocess production cross sections for color-singlet technipions are
listed for some simple models in Section~3. The most promising processes
involve production of an isovector technirho $\tro$ resonance and its
subsequent decay into technipion pairs. Walking technicolor suggests that
$M_{\tro} < 2 M_{\tpi}$, in which case $\tro \ra W_L W_L$ or, more likely,
$W_L \tpi$, where $W_L$ is a longitudinal weak boson. We also discuss a
potentially important new signal: the isoscalar $\omega_T$, degenerate with
$\tro$, and decaying spectacularly to $\gamma \tpi$ and $Z \tpi$. The most
important subprocesses for colored technihadrons are discussed in
Section~4. These involve a color-octet $s$-channel resonance with the same
quantum numbers as the gluon; this technirho $\troct$ dominates colored
technipion pair production. If $M_{\troct} < 2 M_{\tpi}$, then $\troct \ra
\ol q q$ and $gg$, a resonance in dijet production.

The main signatures of topcolor-assisted technicolor, top-pions $\pi_t$ and
the color-octet $V_8$ and singlet $Z'$ of broken topcolor gauge symmetries,
are described in the following report, as are the signatures for quark and
lepton substructure. At the end of the second report, we have provided
a table which summarizes the main processes and sample cross sections at
the Tevatron and LHC. Our reports are not intended to constitute a
complete survey of electroweak and flavor dynamics signatures accessible at
hadron colliders. We have limited our discussion to processes with the
largest production cross sections and most promising signal-to-background
ratios. Even for the processes we list, we have not provided detailed cross
sections for signals and backgrounds. Signal rates depend on masses and
model parameters; they and the backgrounds also depend strongly on detector
capabilities. Experimenters in the detector collaborations will have to
carry out these studies.

\newsec{Overview of Technicolor and Extended Technicolor}

Technicolor---a strong interaction of fermions and gauge bosons at the
scale $\LTC \sim 1\,\tev$---is a scenario for the dynamical breakdown of
electroweak symmetry to electromagnetism
\ref\tcref{S.~Weinberg, Phys.~Rev.~{\bf D19}, 1277 (1979)\semi L.~Susskind,
Phys.~Rev.~{\bf D20}, 2619 (1979).}.
Based on the similar phenomenon of chiral symmetry breakdown in QCD,
technicolor is explicitly defined and completely natural. To account for
the masses of quarks, leptons, and Goldstone ``technipions'' in such a
scheme, technicolor, ordinary color, and flavor symmetries are embedded in
a larger gauge group, called extended technicolor (ETC)
\ref\etc{S.~Dimopoulos and L.~Susskind, Nucl.~Phys.~{\bf B155}, 237
(1979)\semi E.~Eichten and K.~Lane, Phys.~Lett.~{\bf 90B}, 125 (1980).}.
The ETC symmetry is broken down to technicolor and color at a scale $\LETC
= \CO(100\,\tev)$. Many signatures of ETC are expected in the energy regime
of 100~GeV to 1~TeV, the region covered by the Tevatron and Large Hadron
Colliders. For a review of technicolor developments up through 1993, see
Ref.~\ref\tasi{K.~Lane, {\it An Introduction  to Technicolor}, Lectures
given at the 1993 Theoretical Advanced Studies Institute, University of
Colorado, Boulder, published in ``The Building Blocks of Creation'', edited
by S.~Raby and T.~Walker, p.~381, World Scientific (1994).}.

The principal signals in hadron collider experiments of ``classical''
technicolor and extended technicolor were discussed in
Ref.~\ref\ehlq{E.~Eichten, I.~Hinchliffe, K.~Lane and C.~Quigg,
Rev.~Mod.~Phys.~{\bf 56}, 579 (1984); Phys.~Rev.~{\bf D34}, 1547 (1986).}.
In the minimal technicolor model, containing just one technifermion
doublet, the only prominent signals in high energy collider experiments are
the modest enhancements in longitudinally-polarized weak boson production.
These are the $s$-channel color-singlet technirho resonances near
1.5--2~TeV: $\troz \ra W_L^+W_L^-$ and $\tropm \ra W_L^\pm Z_L^0$. The
small $O(\alpha^2)$ cross sections of these processes and the difficulty of
reconstructing weak-boson pairs with reasonable efficiency make observing
these enhancements a challenge. Nonminimal technicolor models are much more
accessible because they have a rich spectrum of lower energy technirho
vector mesons and technipion ($\tpi$) states into which they may decay. In
the one-family model, containing one isodoublet each of color-triplet
techniquarks $(U,D)$ and color-singlet technileptons $(N,E)$, the
technifermion chiral symmetry is $SU(8) \otimes SU(8)$. There are~63
$\rho_T$ and $\tpi$, classified according to how they transform under
ordinary color $SU(3)$ times weak isospin $SU(2)$. The technipions are
$\tpipr \in (1,1)$; $W^\pm_L, Z^0_L$ and $\tpipm, \tpiz \in (1,3)$;  color
octets $\etat \in (8,1)$ and $\octpipm, \octpiz \in (8,3)$; and
color-triplet leptoquarks $\tpiql,\ts \tpilq \in (3,3) \oplus (3,1) \oplus
(\ol 3,3)\oplus(\ol 3,1)$. The $\rho_T$ belong to the same representations.

Because of the conflict between constraints on flavor-changing neutral
currents and the magnitude of ETC-generated quark, lepton and technipion
masses, classical technicolor was superseded a decade ago by ``walking''
technicolor. In this kind of gauge theory, the strong technicolor coupling
$\atc$ runs very slowly for a large range of momenta, possibly all the way
up to the ETC scale---which must be several 100~TeV to suppress FCNC. This
slowly-running coupling permits quark and lepton masses as large as
a few~GeV to be generated from ETC interactions at this very high scale
\ref\wtc{B.~Holdom, Phys.~Rev.~{\bf D24}, 1441 (1981);
Phys.~Lett.~{\bf 150B}, 301 (1985)\semi
T.~Appelquist, D.~Karabali and L.~C.~R. Wijewardhana,
Phys.~Rev.~Lett.~{\bf 57}, 957 (1986);
T.~Appelquist and L.~C.~R.~Wijewardhana, Phys.~Rev.~{\bf D36}, 568
(1987)\semi 
K.~Yamawaki, M.~Bando and K.~Matumoto, Phys.~Rev.~Lett.~{\bf 56}, 1335
(1986) \semi
T.~Akiba and T.~Yanagida, Phys.~Lett.~{\bf 169B}, 432 (1986).}.

Walking technicolor models require a large number of technifermions in
order that $\atc$ runs slowly. These fermions may belong to many copies of
the fundamental representation of the technicolor gauge group, to a few
higher dimensional representations, or to both. This fact inspired a new
kind of model, ``multiscale technicolor'', and a very different
phenomenology
\ref\multi{K. Lane and E. Eichten, Phys. Lett. {\bf B222}, 274 (1989)\semi
K.~Lane and M.~V.~Ramana, Phys.~Rev.~{\bf D44}, 2678 (1991).}.
In multiscale models, there typically are two widely separated scales of
electroweak symmetry breaking, with the upper scale set by the weak decay
constant $F_\pi = 246\,\gev$. Technihadrons associated with the lower scale
may be so light that they are within reach of the Tevatron collider; they
certainly are readily produced {\it and detected} at the LHC. An important
consequence of walking technicolor is that technipion masses are enhanced
so that $\rho_T \ra \tpi\tpi$ decay channels may be closed. If this
happens, then $\tro \ra W_L W_L$ or $W_L \tpi$ and $\troct \ra$~dijets. If
the $\tpi\tpi$ channels are open, they are resonantly produced at large
rates---of order 10~pb at the Tevatron and several nanobarns at the
LHC---and, given the recent successes and coming advances in heavy flavor
detection, many of these technipions should be reconstructable in the
hadron collider environment.

Another major advance in technicolor came in the past two years with the
discovery of the top quark
\ref\toprefs{F.~Abe, et al., The CDF Collaboration, Phys.~Rev.~Lett.~{\bf
73}, 225 (1994); Phys.~Rev.~{\bf D50}, 2966 (1994); Phys.~Rev.~Lett.~{\bf
74}, 2626 (1995) \semi
S.~Abachi, et al., The D\O\ Collaboration, Phys.~Rev.~Lett.~{\bf
74}, 2632 (1995).}.
Theorists have concluded that ETC models cannot explain the top quark's
large mass without running afoul of either cherished notions of naturalness
or experimental constraints from the $\rho$ parameter and the $Z \ra \ol b
b$ decay rate
\ref\zbbexp{A.~Blondel, Rapporteur talk at the International Conference
on High Energy Physics, Warsaw (July 1996)},
\ref\zbbth{R.~S.~Chivukula, S.~B.~Selipsky, and E.~H.~Simmons,
Phys.~Rev.~Lett.~{\bf 69} 575, (1992)\semi
R.~S.~Chivukula, E.~H.~Simmons, and J.~Terning,
Phys.~Lett.~{\bf B331} 383, (1994), and references therein.}.
This state of affairs has led to ``topcolor-assisted technicolor'' (TC2).
In TC2, as in top-condensate models of electroweak symmetry breaking
\ref\topcondref{Y.~Nambu, in {\it New Theories in Physics}, Proceedings of
the XI International Symposium on Elementary Particle Physics, Kazimierz,
Poland, 1988, edited by Z.~Adjuk, S.~Pokorski and A.~Trautmann (World
Scientific, Singapore, 1989); Enrico Fermi Institute Report EFI~89-08
(unpublished)\semi
V.~A.~Miransky, M.~Tanabashi and K.~Yamawaki, Phys.~Lett.~{\bf
221B}, 177 (1989); Mod.~Phys.~Lett.~{\bf A4}, 1043 (1989)\semi
W.~A.~Bardeen, C.~T.~Hill and M.~Lindner, Phys.~Rev.~{\bf D41},
1647 (1990).},
\ref\topcref{C.~T. Hill, Phys.~Lett.~{\bf 266B}, 419 (1991) \semi
S.~P.~Martin, Phys.~Rev.~{\bf D45}, 4283 (1992);
{\it ibid}~{\bf D46}, 2197 (1992); Nucl.~Phys.~{\bf B398}, 359 (1993);
M.~Lindner and D.~Ross, Nucl.~Phys.~{\bf  B370}, 30 (1992)\semi
R.~B\"{o}nisch, Phys.~Lett.~{\bf 268B}, 394 (1991)\semi
C.~T.~Hill, D.~Kennedy, T.~Onogi, H.~L.~Yu, Phys.~Rev.~{\bf D47}, 2940 
(1993).},
almost all of the top quark mass arises from a new strong ``topcolor''
interaction. To maintain electroweak symmetry between (left-handed)
top and bottom quarks
and yet not generate $m_b \simeq m_t$, the topcolor gauge group is
generally taken to be $SU(3)\otimes U(1)$, with the $U(1)$ providing the
difference between top and bottom quarks. Then, in order that topcolor
interactions be natural---i.e., that their energy scale not be far above
$m_t$---and yet not introduce large weak isospin violation, it is necessary
that electroweak symmetry breaking is still due mainly to technicolor
interactions
\ref\tctwohill{C.~T.~Hill, Phys.~Lett.~{\bf 345B}, 483 (1995).}.
In TC2 models, ETC interactions are still needed to generate the light and
bottom quark masses, contribute a few~GeV to $m_t$, and give mass
to the technipions. The scale of ETC interactions still must be hundreds
of~TeV to suppress FCNC and, so, the technicolor coupling must still walk.
Two recent papers developing the TC2 scenario are in
Ref.~\ref\tctwoklee{K.~Lane and E.~Eichten, Phys.~Lett.~{\bf B352}, 382
(1995) \semi
K.~Lane, Phys.~Rev.~{\bf D54}, 2204 (1996).}.
Although the phenomenology of TC2 is in its infancy, it is expected to
share general features with multiscale technicolor: many technihadron
states, some carrying ordinary color, some within range of the Tevatron,
and almost all easily produced and detected at the LHC at moderate
luminosities.

We assume throughout that the technicolor gauge group is $\sutc$ and that
its gauge coupling walks. A minimal, one-doublet model can have a walking
$\atc$ only if the technifermions belong to a large non-fundamental
representation. For nonminimal models, we generally consider the
phenomenology of the lighter technifermions transforming according to the
fundamental~($\Ntc$) representation; some of these may also be ordinary
color triplets. In almost all respects, walking models are very different
from QCD with a few fundamental $SU(3)$ representations. Thus, arguments
based on naive scaling from QCD and on large-$\Ntc$ certainly are suspect.
In TC2, there is no need for large isospin splitting in the technifermion
sector associated with the top-bottom mass difference. This simplifies our
discussion greatly.


\newsec{Color-Singlet Technipion Production}

The $\tro \ra W^+W^-$ and $W^\pm Z^0$ signatures of the minimal model were
discussed in Ref.~\ehlq. The principal change due to the large
representation and walking is that scaling the $\tro \ra \tpi\tpi$ coupling
$\atro$ from QCD is questionable. It may be smaller than usually assumed
and lead to a narrower $\tro$. There is also the possibility that, because
of its large mass (naively, 1.5--2~TeV), the $\tro$ has a sizable branching
ratio to four-weak-boson final states. To our knowledge, neither of these
possibilities has been investigated.

From now on, we consider only nonminimal models which, we believe, are much
more likely to lead to a satisfactory walking model. They have a rich
phenomenology with many diverse, relatively accessible signals. The masses
of technipions in these models arise from broken ETC and ordinary color
interactions. In walking models we have studied, they lie in the range
100--600~GeV; technirho vector meson masses are expected to lie between 200
and 1000~GeV (see, e.g., Ref.~\multi).

Color-singlet technipions, including longitudinal weak bosons $W_L$ and
$Z_L$, are pair-produced via the Drell-Yan process in hadron collisions.
Their $\CO(\alpha^2)$ production rates at the Tevatron and LHC are probably
unobservably small compared to backgrounds {\it unless} there are fairly
strong color-singlet technirho resonances not far above threshold. To
parameterize the cross sections simply, we consider a model containing two
isotriplets of technipions which mix $W_L^\pm$, $Z_L^0$ with a triplet of
mass-eigenstate technipions $\tpi^{\pm,0}$~\multi,
\ref\tctpi{E.~Eichten and K.~Lane, ``Low-Scale Technicolor at the
Tevatron'', FERMILAB-PUB-96/075-T, BUHEP-96-9, hep-ph/9607213; to appear in
Physics Letters~B.}.
We assume that the lighter isotriplet $\tro$ decays into pairs of the state
$\vert\Pi_T\rangle = \sin\chi \ts \vert W_L\rangle + \cos\chi \ts
\vert\tpi\rangle$, leading to the processes
\eqn\singlet{\eqalign{
q \ol q' \ra W^\pm \ra \tropm &\ra \ts\ts W_L^\pm Z_L^0; \quad W_L^\pm
\tpiz, \ts\ts \tpipm Z_L^0; \quad \tpipm \tpiz \cr\cr
q \ol q \ra \gamma, Z^0 \ra \troz &\ra \ts\ts W_L^+ W_L^-; \quad W_L^\pm
\tpimp; \quad \tpip \tpim \ts. \cr}}
The $s$-dependent $\tro$ partial widths are given by (assuming no other
channels, such as colored techipion pairs, are open)
\eqn\singwidth{
\Gamma(\tro \ra \pi_A \pi_B;s) = {2 \atro \CC^2_{AB}\over{3}} \ts
{\ts\ts p_{AB}^3\over {s}} \ts,}
where $p_{AB}$ is the technipion momentum and $\CC^2_{AB} = \sin^4\chi$,
$2\sin^2\chi \cos^2\chi$, $\cos^4\chi$ for $\pi_A \pi_B = W_L W_L$,
$W_L\tpi+\tpi W_L$, $\tpi\tpi$, respectively. The $\tro \ra \tpi\tpi$
coupling $\atro$ obtained by naive scaling from QCD is~\ehlq
\eqn\alpharho{\atro = 2.91 \left({3\over{\Ntc}}\right)\ts.}

Technipion decays are mainly induced by ETC interactions which couple them
to quarks and leptons. These couplings are Higgs-like, and so technipions
are expected to decay into heavy fermion pairs:
\eqn\singdecay{\eqalign{
\tpiz &\ra \cases{b \ol b &if $M_{\tpi} < 2 m_t$,  \cr
t \ol t &if $M_{\tpi} > 2 m_t$; \cr} \cr
\tpip &\ra \cases{c \ol b \ts\ts\ts {\rm or} \ts\ts\ts c \ol s, \ts\ts
\tau^+ \nu_\tau 
&if $M_{\tpi} < m_t + m_b$, \cr
t \ol b &if $M_{\tpi} > m_t + m_b$. \cr} \cr}}
An important caveat to this rule applies to TC2 models. There, only a
few~GeV of the top mass arises from ETC interactions. Then, the $b \ol b$
mode competes with $t \ol t$ for $\tpiz$; $c \ol b$ or $c \ol s$ compete
with $t \ol b$ for $\tpip$. Note that, since the decay $t \ra \tpip b$ is
strongly suppressed in TC2 models, the $\tpip$ can be much lighter than the
top quark.

The $\tro\ra \pi_A \pi_B$ cross sections are well-approximated by
\eqn\singcross{
{d\hat\sigma(q_i \ol q_j \ra \tro^{\pm,0} \ra \pi_A\pi_B) \over{dz}} = 
{\pi \alpha^2 p_{AB}^3 \over{3 \shat^{5/2}}}
\ts {M^4_{\tro} \ts \ts (1-z^2) \over
{(\shat - M_{\tro}^2)^2 + \shat \Gamma_{\tro}^2}} \ts A_{ij}^{\pm,0}(\shat)
\CC^2_{AB} \ts,}
where $\shat$ is the subprocess energy, $z = \cos\theta$ is the
$\pi_A$ production angle, and $\Gamma_{\tro}$ is the $\shat$-dependent
total width of $\tro$. Ignoring Kobayashi-Maskawa mixing angles, the
factors $A_{ij}^{\pm,0} = \delta_{ij} A^{\pm,0}$ are
\eqn\bfactors{\eqalign{
A^\pm &= {1 \over {4 \sin^4\thw}} \biggl({\shat \over {\shat -
M_W^2}}\biggr)^2 \cr
A^0   &= \biggl[Q_i + {2 \cos 2\thw \over {\sin^2 2\thw}} \ts
(T_{3i} - Q_i \sin^2\thw) \biggl({\shat \over {\shat - M_Z^2}}\biggr)
\biggl]^2 \cr 
&\ts + \biggl[Q_i - {2 Q_i \cos 2\thw \sin^2\thw \over{\sin^2
2\thw}} \ts \biggl({\shat \over {\shat - M_Z^2}}\biggr) \biggl]^2 \ts.}}
Here, $Q_i$ and $T_{3i}$ are the electric charge and third component of
weak isospin for~$q_{i L,R}$. Production rates of several picobarns increase
by factors of 5--10 at the LHC; see the table in the following Report~II.

If the isospin of technifermions is approximately conserved, there is an
isoscalar partner $\omega_T$ of the $\tro$ that is nearly degenerate with
it and may be produced at a comparable rate.  The walking technicolor
enhancement of technipion masses almost certainly closes off the
isospin-conserving decay $\omega_T \ra \Pi^+_T \Pi^-_T \Pi^0_T$. Even the
triply-suppressed mode $W^+_L W^-_L Z_L$ has little or no phase space for
$M_{\omega_T} \simle 300\,\gev$. Thus, we may expect the main decays to be
$\omega_T \ra \gamma \Pi_T^0$, $Z \Pi_T^0$, and $\Pi_T^+ \Pi_T^-$. In terms
of  mass eigenstates, these modes are $\omega_T \ra \gamma \tpiz$, $\gamma
Z_L$, $Z \tpiz$, $Z Z_L$; $\gamma \tpipr$, $Z \tpipr$; and $W^+_L W^-_L$,
$\tpipm W^\mp_L$, $\tpip\tpim$.\foot{The modes $\omega_T \ra \gamma Z_L$,
$Z Z_L$ were considered for a one-doublet technicolor model in
Ref.~\ref\cg{R.~S.~Chivukula and M.~Golden, Phys.~Rev.~{\bf D41}, 2795
(1990).}. We have estimated the branching ratios for the isospin-violating
decays $\tro \ra \gamma \tpiz$, $Z \tpiz$ and found them to be negligible
unless the mixing angle $\chi$ is very small.} It is not possible to
estimate the relative magnitudes of the decay amplitudes without an
explicit model of the $\omega_T$'s constituent technifermions. Judging from
the decays of the ordinary $\omega$, we expect $\omega_T \ra \gamma \tpiz
(\tpipr)$, $Z \tpiz (\tpipr)$ to dominate, with the former mode favored by
phase space.

The $\omega_T$ is produced in hadron collisions just as the $\tro^0$, via
its vector-meson-dominance coupling to $\gamma$ and $Z^0$. For
$M_{\omega_T} \simeq M_{\tro}$, the $\omega_T$ production cross section
should be approximately $|Q_U + Q_D|^2$ times the $\tro^0$ rate, where
$Q_{U,D}$ are the electric charges of the $\omega_T$'s constituent
technifermions. The principal signatures for $\omega_T$ production, then,
are $\gamma + \ol b b$ and $\ell^+\ell^-$ (or $\nu \ol \nu$) $+ b \ol b$,
with $M_{\ol b b} = M_{\tpi}$.

In the one-family and other models containing colored as well as
color-singlet technifermions, there are singlet and octet technipions
that are electroweak isosinglets commonly denoted $\tpipr$ and $\etat$.
These are singly-produced in gluon fusion. Depending on the technipion's
mass, it is expected to decay to $\ol b b$ (and, possibly, $gg$) or to $\ol
t t$~\ehlq,
\ref\etatrefs{E.~Farhi and L.~Susskind Phys.~Rev.~{\bf D20}
(1979)~3404\semi
S.~Dimopoulos, Nucl.~Phys.~{\bf B168} (1980)~69 \semi
T.~Appelquist and G.~Triantaphyllou, Phys.~Rev.~Lett.~{\bf
69},2750 (1992) \semi
T.~Appelquist and J.~Terning, Phys.~Rev.~{\bf D50}, 2116 (1994)\semi
E.~Eichten and K.~Lane, Phys.~Lett.~{\bf B327}, 129 (1994)\semi
K.~Lane, Phys.~Rev.~{\bf D52}, 1546 (1995).}.
With $\Pi^0 = \tpipr$ or $\etat$, and with constituent technifermions
transforming according to the $\Ntc$~representation of $\sutc$,
the decay rates are
\eqn\Piwidths{\eqalign{
\Gamma(\Pi^0 \ra gg) &= {\CC_\Pi \aqcd^2 \ts \Ntc^2 \ts M_\Pi^3  \over {128
\ts \pi^3 \ts F_T^2}} \ts, \cr\cr
\Gamma(\Pi^0 \ra \ol q q) &= {\gamma_q^2 \ts m_q^2 \ts M_\Pi \ts \beta_q
\over {16 \pi F_T^2}} \ts.\cr}}
Here, $\beta_q = \sqrt{1 - 4m^2_q/M_\Pi^2}$ is the quark velocity. The
$SU(3)$-color factor $\CC_\Pi$ is determined by the triangle-anomaly graph
for $\Pi^0 \ra gg$. In the one-family model, $\CC_\Pi = {\tx{4\over{3}}}$ for
the singlet $\tpipr$ and ${\tx{5\over{3}}}$ for the octet $\etat$; values of
$\CO(1)$ are expected in other models. The technipion decay constant $F_T$
is discussed below. The dimensionless factor $\gamma_q$ allows for model
dependence in the technipions' couplings to $\ol q q$. In classical ETC
models, we expect $|\gamma_q| = \CO(1)$. In TC2 models, $|\gamma_q| =
\CO(1)$ for the light quarks and, possibly, the $b$-quark, but $|\gamma_t|
= \CO({\rm few} \ts \gev/m_t) \ll 1$; there will be no $\etat$ enhancement
of $\ol t t$ production in topcolor-assisted technicolor.

The gluon fusion cross section for production and decay of $\Pi^0$ to heavy
$\ol q q$ is isotropic:
\eqn\sigPi{
{d \hat \sigma(gg \ra \Pi^0 \ra \ol q q) \over {d z}} =
{\pi \CN_C \over{32}} \ts {\Gamma(\Pi^0 \ra gg)\ts \Gamma(\Pi^0 \ra \ol q q)
\over {(\shat - M_\Pi^2)^2 + \shat \ts \Gamma^2_{\Pi^0} }} \ts,}
where $\CN_C = 1$ (8) for $\tpipr$ ($\etat$). The decay rates
and cross sections are contolled by the technipion decay constant
$F_T$. In the standard one-family model, $F_T = 123\,\gev$ and the
enhancements in $\ol q q$ production are never large enough to see above
background (unless $\Ntc$ is unreasonably large). In multiscale models and,
we expect, in TC2 models, $F_T$ may be considerably smaller. For example,
in the multiscale model considered in Ref.~\multi, $F_T =30$--$50\,\gev$;
in the TC2 model of Ref.~\tctwoklee, $F_T = 80\,\gev$. Since the total
hadronic cross section,
\eqn\narrowPi{
\sigma(p p^\pm \ra \Pi^0 \ra \ol q q) \simeq {\pi^2 \over {2s}} \ts
{\Gamma(\Pi^0 \ra gg) \ts \Gamma(\Pi^0 \ra \ol q q) \over {M_\Pi \ts
\Gamma_{\Pi^0}}}
\ts \int d \eta_B \ts f_g^p\biggl({M_\Pi\over{\sqrt{s}}}
e^{\eta_B}\biggr)
\ts f_g^p\biggl({M_\Pi\over{\sqrt{s}}} e^{-\eta_B}\biggr) \ts,}
scales as $1/F_T^2$, small decay constants may lead to observable
enhancements of $\ol t t$ production in standard multiscale technicolor and
in $\ol b b$ production in TC2. Sample rates are given in the table in
Report~II.

In models containing colored technifermions, color-singlet technipions are
also pair-produced in the isospin $I=0$ channel via gluon fusion. This
process involves intermediate states of color-triplet and octet
technipions. Again, the subprocess cross section is isotropic; it is given by
\ref\tpitev{K.~Lane, Phys.~Lett.~{\bf B357}, 624 (1995)\semi also see
T.~Lee, Talk presented at International Symposium on Particle Theory and
Phenomenology, Ames, IA, May 22-24, 1995, FERMILAB-CONF-96-019-T,
hep-ph/9601304, (1996).}
\eqn\dsggpp{\eqalign{
& {d \hat\sigma(gg \ra \tpip \tpim) \over{dz}} =
2{d \hat\sigma(gg \ra \tpiz \tpiz) \over{dz}} \cr
& \qquad = {\aqcd^2 \beta \over {2^{15} \pi^3 F_T^4 \shat}}
\ts \biggl\vert T(R) \ts \left[C_R \ts \left(\shat  - \twothirds(2M_R^2
+ M_{\pi_T}^2)\right) + D_R \right] \ts \left(1 +
2\CI(M_R^2,\shat)\right)\biggl\vert^2 \ts. \cr }}
Here, $\beta = 2p/\rshat$ is the technipion velocity.
The sum is over $SU(3)$ representations $R =3,8$ of the $\tpi$ and $T(R)$
is the trace of the square of their $\suc$-generator matrices: $T(R) =
\half$ for triplets (dimension $d(R) = 3$), 3 for octets ($d(R) = 8$). The
factors $C_R$ and $D_R$ for the one-family model and a multiscale model
are:

\medskip

\centerline{\vbox{\offinterlineskip
\hrule\hrule
\halign{&\vrule#&
  \strut\quad#\hfil\quad\cr\cr
height4pt&\omit&&\omit&&\omit&&\omit&&\omit&\cr\cr
&\hfill Model \hfill&&\hfill $C_3$ \hfill&&\hfill
$C_8$  \hfill&&\hfill $D_3$\hfill&&\hfill $D_8$\hfill &\cr\cr
height4pt&\omit&&\omit&&\omit&&\omit&&\omit&\cr\cr
\noalign{\hrule\hrule}
height4pt&\omit&&\omit&&\omit&&\omit&&\omit&\cr\cr
&One--Family $\tpi\tpi$&&\hfill${10\over{3}}$\hfill&&\hfill${1\over{3}}$
\hfill&&
\hfill${16\over {9}} M_3^2$\hfill&&\hfill ${4\over{9}}
M_8^2$\hfill&\cr\cr
\noalign{\hrule}
height4pt&\omit&&\omit&&\omit&&\omit&&\omit&\cr\cr
&Multiscale $\pi_{\ol Q Q} \pi_{\ol Q Q}$
&&\hfill${8\over{3}}$\hfill&&\hfill${4\over{3}}$\hfill&&  
\hfill${32\over {9}} M_3^2$\hfill&&\hfill ${16\over{9}}
M_8^2$\hfill&\cr\cr
\noalign{\hrule}
height4pt&\omit&&\omit&&\omit&&\omit&&\omit&\cr\cr
&Multiscale $\pi_{\ol L L} \pi_{\ol L L}$
&&\hfill$8$\hfill&&\hfill$0$\hfill&&  
\hfill${16\over {3}}(2M_{\pi_T}^2 - M_3^2)$\hfill&&\hfill
$0$\hfill&\cr\cr
height4pt&\omit&&\omit&&\omit&&\omit&&\omit&\cr\cr}
\hrule\hrule}}
\medskip

\noindent The integral~$\CI$ is
\eqn\zint{\eqalign{
\CI(M^2,s) &\equiv  \int_0^1 dx \ts dy \ts {M^2 \over {xys - M^2 +
i\epsilon}} \ts \theta(1-x-y) \cr
&=\cases{-M^2 /2s \left[ \pi - 2 \arctan \sqrt{4 M^2/s -1}
\right]^2 & for $s <  4M^2$ \cr
M^2/2s \left[ \ln \left({1 + \sqrt{1 - 4 M^2/s} \over
{1 - \sqrt{1 - 4 M^2/s}}}\right) - i\pi\right]^2  &for $s > 4M^2$ \ts.}
\cr}}
The rates at the Tevatron are at most comparable to those enhanced by
technirhos; they are considerably greater at the LHC because the fusing
gluons are at low~$x$ (see the table in Report~II). An interesting feature
of this cross section is that the $\tpi\tpi$ invariant mass distribution
peaks near the color-triplet and octet technipion thresholds, which can be
well above $2 M_{\tpi}$. It is possible that mixed modes such as $W_L^\pm
\tpimp$ and $Z_L \tpiz$ are also produced by gluon fusion, with the rates
involving mixing angles such as $\chi$ in Eq.~\singcross.

\newsec{Color-Octet Technirho Production and Decay to Jets and Technipions}

Models with an electroweak doublet of color-triplet techniquarks $(U,D)$
have an octet of $I=0$ technirhos, $\troct$, with the same quantum numbers
as the gluon. The $\troct$ is produced strongly in $\ol q q$ and $gg$
collisions. Assuming the one-family model for simplicity, there are the 63
technipions listed in Section~2. The color-singlet and octet technipions
decay as in Eq.~\singdecay\ above. The leptoquark decay modes are expected
to be
\eqn\leptoquark{\eqalign{
\tpiun &\ra \cases{c \ol \nu_\tau &if $M_{\tpi} <  m_t$,  \cr
t \ol \nu_\tau &if $M_{\tpi} > m_t$; \cr} \cr
\tpiue &\ra \cases{c \tau^+ &if $M_{\tpi} <  m_t$,  \cr
t \tau^+ &if $M_{\tpi} > m_t$; \cr} \cr
\tpidn &\ra b \ol \nu_\tau \ts; \cr
\tpide &\ra b \tau^+ \ts.  \cr}}
The caveat regarding technipion decays to top quarks in TC2 models still
applies.

There are two possibilities for $\troct$ decays~\multi. If walking
technicolor enhancements of the technipion masses close off the $\tpi\tpi$
channels, then $\troct \ra \ol q q,\ts gg \ra \jets$. The color-averaged
$\CO(\aqcd^2)$ cross sections are given by
\eqn\dsjets{\eqalign{
& {d\hat \sigma(\ol q_i q_i \ra \ol q_i q_i) \over {d z}} = 
{2 \pi \aqcd^2 \over {9 \shat}} \left\{ \ts \bigl| \CDgg(\shat) \bigr|^2 \ts
\left({\uhat^2 + \that^2 \over {\shat^2}} \right)
- \twothirds \ts {\rm Re}\ts \CDgg(\shat) \ts \left({\uhat^2 \over
{\shat\that}}\right)
+ {\shat^2 + \uhat^2 \over {\that^2}} \right\}
\ts; \cr\cr
& {d\hat \sigma(\ol q_i q_i \ra \ol q_j q_j)
\over {dz}} = {2 \pi \aqcd^2 \over {9 \shat}} \ts \bigl| \CDgg(\shat)
\bigr|^2 \ts \left({\uhat^2 + \that^2 \over {\shat^2}} \right)
\ts; \cr\cr
& {d\hat \sigma(\ol q_i q_i \ra gg) \over {d z}} =
{64 \over {9 }} {d\hat \sigma(gg \ra q_i \ol q_i) \over {d z}} 
= {4 \pi \aqcd^2 \over {3 \shat}}
\left\{ \ts \bigl| \CDgg(\shat) - 1 \bigr|^2 \ts
{2\uhat\that \over{\shat^2}}
+ {\tx{{4 \over {9}}}} \biggl({\uhat \over {\that}} + {\that \over {\uhat}}
\biggr) 
- {\uhat^2 + \that^2 \over {\shat^2}} \right\}
\ts; \cr\cr
& {d\hat \sigma(gg \ra gg) \over {d z}} = {9 \pi \aqcd^2 \over {4 \shat}}
\biggl\{ \ts 3 - {\uhat\that\over{\shat^2}} - {\that\shat\over{\uhat^2}} 
- {\shat\uhat\over{\that^2}} \cr
& \qquad\qquad\qquad + {\tx{{1 \over {4}}}} \bigl| \CDgg(\shat) - 1 \bigr|^2
\ts\left({\uhat - \that \over{\shat}}\right)^2
- {\tx{{1 \over {4}}}} {\rm Re}(\CDgg(\shat) -1) \ts 
{\left(\uhat - \that\right)^2\over {\uhat\that}}
\biggr\}
\ts; \cr\cr
& {d\hat \sigma(q_i q_j \ra q_i q_j) \over {d z}} =
{d\hat \sigma(\ol q_i \ol q_j \ra \ol q_i \ol q_j) \over {d z}}
= {d\hat \sigma(q_i \ol q_j \ra q_i \ol q_j) \over {d z}}
= {2 \pi \aqcd^2 \over {9 \shat}} \ts
\left({\shat^2 + \uhat^2 \over {\that^2}}\right)
\ts; \cr\cr
& {d\hat \sigma(q_i q_i \ra q_i q_i) \over {d z}} =
{d\hat \sigma(\ol q_i \ol q_i \ra \ol q_i \ol q_i) \over {d z}}
= {2 \pi \aqcd^2 \over {9 \shat}} \ts \left\{
{\shat^2 + \uhat^2 \over{\that^2}} + {\that^2 + \uhat^2 \over{\shat^2}}
- \twothirds {\shat^2 \over {\uhat\that}} \right\} \ts;\cr \cr
& {d\hat \sigma(g q_i \ra g q_i) \over {d z}} =
{d\hat \sigma(g \ol q_i \ra g \ol q_i) \over {d z}}
= {\pi \aqcd^2 \over {2 \shat}} \ts (\shat^2 + \uhat^2) \ts
\left( \ts {1 \over {\that^2}} - {4 \over {9 \shat\uhat}} \right)
\ts . \cr}}
Here, $z = \cos\theta$, $\that = -\half \shat(1 - z)$, $\uhat = -\half
\shat(1 + z)$ and it is understood that $q_i \ne q_j = u,d,c,s,b$
contribute to dijet events. Only the $s$-channel gluon propagator was
modified to include the $\troct$ resonance. Here and below, we use the
dimensionless propagator factors $\CDgg$ and $\CDgrho$
\eqn\cdg{\eqalign{
\CDgg(s) &= {s - M_{\troct}^2 + i \ecm \ts \Gamma_{\troct}(s) \over
{s(1- \aqcd(s)/\atro) - M_{\troct}^2 + i \ecm \ts \Gamma_{\troct}(s)}}
\ts, \cr\cr
\CDgrho(s) &= {s \over
{s(1- \aqcd(s)/\atro) - M_{\troct}^2 + i \ecm \ts
\Gamma_{\troct}(s)}} \ts. \cr}}
If $M_{\troct} < 2 M_{\tpi}$, the $s$-dependent $\troct$ width is the sum
of (allowing for multijet $\ol t t$ final states, assumed light compared to
$\sqrt{s}$)
\eqn\troctwidth{\eqalign{
&\sum_{i=1}^6 \Gamma(\troct \ra \ol q_i q_i) = {6 \over {3}} {\aqcd^2(s)
\over {\atro}} \ecm \ts , \cr
&\Gamma(\troct \ra gg) = {\aqcd^2(s) \over {\atro}} \ecm \ts . \cr }}
A search for the dijet signal of $\troct$ has been carried out by the CDF
Collaboration; see Ref.~\ref\cdfdijet{F.~Abe, et al., The CDF
Collaboration, Phys.~Rev.~Lett.~{\bf 74}, 3538 (1995).}
for a detailed discussion of expected signal and background rates.
Rough signal-to-background estimates are given in the table in
Report~II. They are sizable at the Tevatron and LHC, but are sensitive to
jet energy resolutions.

Colored technipions are pair-produced in hadron collisions through
quark-antiquark annihilation and gluon fusion. If the $\troct \ra
\tpi\tpi$ decay channels are open, this production is resonantly
enhanced. The subprocess cross sections, averaged over initial colors
and summed over the colors $B$, $C$ of technipions, are given by
\eqn\qbqpipi{\sum_{B,C} {d\hat \sigma(\ol q_i q_i \ra \pi_B \pi_C) \over
{dz}} = {\pi \aqcd^2(\shat) \beta^3 \over {9 \shat}} \ts \CS_{\pi} T(R) \ts
\bigl(1 - z^2 \bigr) \ts \bigl| \CDgg + \CDgrho \bigr|^2 \ts,}
\eqn\ggpipi{\eqalign{
&\sum_{B,C}{d\hat \sigma(gg \ra \pi_B \pi_C)\over {dz}} =
 {\pi \aqcd^2(\shat) \beta \over {\shat}} \ts
\CS_{\pi} T(R) \biggl\{{3 \over {32}} \ts \beta^2 \ts z^2 \ts
\biggl[\bigl| \CDgg + \CDgrho \bigr|^2 \cr
&\qquad\qquad -{2 \beta^2 \ts (1-z^2) \over {1-\beta^2 z^2}}
\ts {\rm Re}\ts \left(\CDgg + \CDgrho \right)
+ 2 \biggl({\beta^2 \ts (1-z^2)
\over {1 - \beta^2 z^2}} \biggr)^2 \biggr] \cr
&\qquad \qquad + \left({T(R) \over {d(R)}} - {3 \over {32}} \right)
\biggl[ {(1 - \beta^2)^2 + \beta^4 \ts (1-z^2)^2
\over {(1 - \beta^2 z^2)^2}} \biggr] \biggr\}
\ts,\cr}}
where $\beta$ is the technipion velocity and $z = \cos\theta$.
The symmetry factor $\CS_{\pi} = 1$ for each channel of $\tpilq \tpiql$ and
for $\octpip \octpim$; $\CS_{\pi} = \half$ for the identical-particle final
states, $\octpiz \octpiz$ and $\etat\etat$. The $SU(3)$ group factors
$T(R)$ and $d(R)$ for $R=3,8$ were defined above at Eq.~\dsggpp. The
technirho width is now the sum of the $\ol q q$ and $gg$ partial widths and
\eqn\octwidth{
\sum_{B,C} \Gamma(\tro \ra \pi_B \pi_C;s) = {\atro \CS_\pi T(R) \over {3}}
{\ts\ts\ts p^3 \over{s}} \ts.}
As indicated in the table in Report~II, pair-production rates for colored
technipions with masses of a few hundred~GeV are several picobarns at the
Tevatron, rising to a few nanobarns at the LHC.

\vfil\eject

\listrefs

\vfil\eject

\bye